\documentclass[sigconf, nonacm]{acmart}

\AtBeginDocument{%
  \providecommand\BibTeX{{%
    \normalfont B\kern-0.5em{\scshape i\kern-0.25em b}\kern-0.8em\TeX}}}

\usepackage{amsmath,amsfonts}

\usepackage{algpseudocode}
\usepackage{algorithm}
\usepackage{graphicx}
\usepackage{textcomp}

\usepackage{microtype} 
\usepackage{booktabs} 
\usepackage{listings}
\usepackage{mathtools}
\usepackage{bm}
\usepackage{enumerate}
\usepackage[framemethod=TikZ]{mdframed}
\usetikzlibrary{shapes.geometric}
\usepackage{xspace}
\usepackage{xcolor}
\usepackage{cleveref}
\usepackage{relsize}
\usepackage[T1]{fontenc} 
\usepackage{color} 
\usepackage{wrapfig}
\usepackage{balance}
\usepackage{multirow}
\usepackage{multicol}
\usepackage[most]{tcolorbox}
\usepackage{array}
\usepackage[normalem]{ulem}
\usepackage{soul}
\usepackage[english]{babel}
\usepackage{subfig}

\definecolor{shadecolor}{RGB}{150,150,150}

\def\HiLi{\leavevmode\rlap{\hbox to \hsize{\color{black!20}\leaders\hrule height .8\baselineskip depth .5ex\hfill}}}

\graphicspath{{./figures/}}

\makeatletter
\newcommand{\algcolor}[2]{%
  \hskip-\ALG@thistlm\colorbox{#1}{\parbox{\dimexpr\linewidth-2\fboxsep}{\hskip\ALG@thistlm\relax #2}}%
}

\makeatother

\newcommand{\setting}[1]{\textsc{\small #1}}
\newcommand{\app}[1]{\textsf{#1}}

\newcommand{\name}{Sonic\xspace}

\newcommand{\bo}{Bayesian optimization\xspace}

\newcommand{\odroid}{Odroid XU4\xspace}
\newcommand{\jetson}{Jetson TX2\xspace}
\newcommand{\xeon}{Intel Xeon Gold\xspace}

\newcommand{\naive}{na\"{\i}ve\xspace}

\newcommand\letequal{\ensuremath{\gets}\xspace}

\makeatletter
\NewDocumentCommand{\LeftComment}{s m}{%
  \Statex \IfBooleanF{#1}{\hspace*{\ALG@thistlm}}\(\triangleright\) #2}
\makeatother

\tcbset{
  frame code={}
  center title,
  left=0pt,
  right=0pt,
  top=0pt,
  bottom=0pt,
  colback=gray!20,
  colframe=white,
  width=\dimexpr\textwidth\relax,
  enlarge left by=0mm,
  boxsep=5pt,
  arc=0pt,outer arc=0pt,
}

\newcolumntype{H}{>{\setbox0=\hbox\bgroup}c<{\egroup}@{}}

\definecolor{lightblue}{rgb}{.90,.95,1}
\definecolor{bestblue}{rgb}{0.078, 0.325, 0.678}
\definecolor{mypink1}{rgb}{0.858, 0.188, 0.478}
\definecolor{mypink3}{cmyk}{0, 0.7808, 0.4429, 0.1412}
\definecolor{mygray}{gray}{0.90}
\definecolor{codehighlight}{rgb}{0.95,0.8,0.8}
\definecolor{codebackground}{rgb}{0.95,0.95,0.95}
\definecolor{lightyellow}{RGB}{255,255,204}

\newenvironment{closeitemize}{\begin{list}{\small $\bullet$}%
    {\usecounter{enumiv} \setlength{\itemsep}{0in} \setlength{\parsep}{0in}
          \setlength{\topsep}{1ex}}
          \def\makelabel##1{\hss\llap{##1}}}%
          {\end{list}}

\begin{document}

\title{\name: A Sampling-based Online Controller for\\Streaming Applications}

\author{Yan Pei}
\email{ypei@cs.utexas.edu}
\affiliation{%
  \institution{University of Texas at Austin}
  \city{Austin}
  \state{TX}
  \country{USA}
}

\author{Keshav Pingali}
\email{pingali@cs.utexas.edu}
\affiliation{%
  \institution{University of Texas at Austin}
  \city{Austin}
  \state{TX}
  \country{USA}
}

\thispagestyle{plain}
\pagestyle{plain}

\begin{abstract}
    \vspace{1pt}
Many applications in important problem domains such as machine learning and computer vision are streaming applications that take a sequence of inputs over time.
It is challenging to find knob settings that optimize the run-time performance of such applications because the optimal knob settings are usually functions of inputs, computing platforms, time as well as user's requirements, which can be very diverse.

Most prior works address this problem by offline profiling followed by training
models for control. However, profiling-based approaches incur large overhead before execution;
it is also difficult to redeploy them in other run-time configurations.

In this paper, we propose \name, a sampling-based online controller for
streaming applications that does not require profiling ahead of time.
Within each phase of a streaming application's execution,
\name utilizes the beginning portion to sample the knob space strategically and aims to pick the optimal knob setting for the rest of the phase, given a user-specified constrained optimization problem.
A hybrid approach of machine learning regressors and \bo are used for better sampling choices.

\name is implemented independent of application, device, input, performance objective and constraints.
We evaluate \name on traditional parallel benchmarks as well as on deep learning
inference benchmarks across multiple platforms. Our experiments show that when
using \name to control knob settings, application run-time performance is only
5.3\% less than if optimal knob settings were used, demonstrating that \name is
able to find near-optimal knob settings under diverse run-time configurations without prior knowledge.

\end{abstract}


\maketitle

\section{Introduction}
\label{sec:intro}

Streaming applications are ubiquitous in important domains such as deep learning, computer vision and media processing.
Those applications are typically long-running applications with inputs supplied
incrementally over time, assuming that the run-time behavior of the application
for successive inputs is stationary. For instance, augmented reality (AR)
applications take in a series of frames to enhance real-world scenes by
computer-generated perceptual information. Optimizing streaming application's
run-time performance is important but challenging because it depends on multiple
factors including inputs, computing devices, time as well as user's requirements.

Most streaming applications have a set of tunable parameters, usually referred
to as \emph{knobs}, that can be adjusted to optimize their performance;
as long as knob values are within certain limits, the correctness of the application is not compromised.
Many streaming applications are executed on parallel platforms, so their performance
is also affected by characteristics of the device such as the number of
cores allocated and their clock frequency.
Finding the optimal device knob setting is challenging because modern compute devices have more and more non-trivial parallelism.

In addition, applications are usually optimized subject to various constraints such as computation accuracy, power/energy consumption, \emph{etc}. For instance, energy consumption is an important factor in lowering the operation cost of data centers as well as lengthen battery life for resource-constrained devices.
Given different combinations of performance objectives and constraints, the
system needs to find different optimal knob settings accordingly.

Furthermore, application's performance is also sensitive to inputs.
For instance, both image size and image content may affect image/video related tasks' performance.
Unlike problems such as n-body simulation~\cite{aarseth2003gravitational} and graph partitioner~\cite{bader2013graph} whose inputs are available at the beginning of the computation for inspection, streaming application's inputs are usually not known before execution and are supplied over time.

The cross-product of applications, devices, inputs, performance objective and constraints makes it very challenging to decipher the optimal application and device knob settings of any run-time configuration. Prior works usually address such problem through offline executions of a set of run-time configurations, or offline profiling, followed by constructing models to predict application's run-time behavior~\cite{capri, capri_sw, poet, neuralvector, thunderx_beacon, rumba, green, caloree, hbm, gmm}.
Models are then used to perform \emph{offline control} or \emph{reactive online
control}. For example, \emph{Capri} builds proxy models for both quality and
running time of a program through offline profiling. These models together with
input features are then used to perform proactive control of approximation~\cite{capri};
\emph{CALOREE} performs online control using estimated performance-energy pareto frontier of one application by matching its behavior with models that have been already profiled on the same machine~\cite{caloree}.

However, there are several drawbacks for such profiling-based approaches.
\begin{closeitemize}
    \item \emph{\textbf{Overhead}}. Profiling incurs large overhead, including compilation, data collection (execution), model training and so on.
        On embedded platforms, profiling takes long time due to limited computation resource while it can be very costly to get profiling data on larger servers.
        Profiling even becomes inapplicable when single execution of an application is expensive, e.g. chemical material design~\cite{griffiths2017constrained}.

    \item \emph{\textbf{Scope}}. Due to the overhead of profiling,
        only a specific subset of the whole run-time configuration space can be profiled.
        For example, prior works usually profile on a small subset of
        inputs and assume it is representative~\cite{capri, rumba};
        Capri~\cite{capri} and Rumba~\cite{rumba} focus on
        optimizing performance under accuracy constraint while
        \emph{POET}~\cite{poet} and CALOREE~\cite{caloree} only target on
        minimizing energy given a performance requirement.
        Users may be interested in other reasonable constrained optimization
        problems such as ``Least resource allocation given a performance requirement'' on super computers.

    \item \emph{\textbf{Portability}}. Models built on profiling data are difficult to be
        ported to other unseen run-time configurations.
        For instance, the relationship between application's performance and power on one device may not hold on new device with distinct hardware architectures.

\end{closeitemize}

While profiling-based approaches tries to anticipate application's run-time behavior through offline data,
we propose to solve such control problem from a pure online control perspective,
in which the optimal knob setting is searched during run time,
given the nature of streaming applications that the run-time behavior for successive inputs is stationary.

Due to complex run-time configuration space,
we formulate the control problem as a \emph{global optimization}
process that searches the optimal knob setting by sampling,
rather than a \emph{traditional feedback control loop} that adjusts knob
settings by comparing quantities of interest with reference values.
In this paper, we propose \emph{\name}, a sampling-based online controller
that does not require profiling ahead of time.
\name consists of two key components: a sampler and a phase detector.

\paragraph{\textbf{Sampler.}}
Within each phase of a streaming application's execution, the sampler utilizes the beginning portion of that phase to sample the knob space and collect the statistics of each sampled knob setting.
After the sampling phase, the controller picks the best knob setting among the
sampled ones based on the performance objective \& constraint of the current
run-time configuration. The picked knob setting will then be used for the following execution.
Since it is challenging to capture and optimize application's run-time behavior
with only limited samples of knob settings, we incorporate combinations of
machine learning regressors and sequential design strategy for global
optimization such as \bo~\cite{bo, bo_tutorial, bo_review} for better sample choices.

\paragraph{\textbf{Phase Detector.}} Streaming application may have distinct phases
during one run due to various reasons such as algorithmic phases, launch of another program on the same device, inputs change, \emph{etc}.
In \name, a phase detector is activated after the sampling period and monitors
the difference between application's current behavior with the recorded
behavior of the picked knob setting during the sampling phase. If the difference is
large, a new sampling period will be activated. In this paper, we assume that
each phase is long enough for the sampling-based online control strategy to be beneficial.

We evaluate \name on the PARSEC parallel benchmark suite~\cite{parsec} and a modified MLPerf inference benchmark suite~\cite{mlperf} across multiple platforms.
The primary contributions of this paper are listed as follows:
\begin{itemize}
    \item To the best of our knowledge, \name is the first sampling-based online controller targeting general streaming applications without dependencies on offline profiling.
    \item \name incorporates a hybrid control approach that consists of sequential design strategies and machine learning regressors to improve the choice of samples.
    \item \name is implemneted independent of applications, devices, inputs, objectives and constraints, so it can be easily ported to different run-time configurations.
        It is also orthogonal to other types of optimizations such as algorithmic improvement, new accelerator and so on.
    \item \name is easy to use. The only extra code needed for application and
        device is an interface to report their performance at run time.
        No need for algorithmic changes such as annotating code, enabling
        re-computation~\cite{green, rumba}, \emph{etc}.
    \item By evaluating on traditional and deep learning benchmarks, our
        experiments show that when using \name, application performance is only
        5.3\% less than if optimal knob settings were used, showing that \name
        is able to find near-optimal knob settings under diverse run-time configurations.
\end{itemize}

\name can be very useful in situations when run-time configurations changes frequently.
For example, devices may be unknown and undeterministic when using cloud
service; user changes input, performance objective/constraint frequently.

\section{Background and Motivation}
\label{sec:bg}

Streaming applications takes a sequence of inputs such as an audio/image sequence, a series of transformations, \emph{etc}. These inputs are usually not available before application starts, and will be supplied over time.
In this paper, we will focus on streaming applications that run on CPUs due to
CPU's advantages such as flexibility and availability over other devices such as
GPUS, ASICs, \emph{etc}.



\subsection{Baseline Performance}
\label{sec:baseline_perf}



In the default run-time setting, application's knobs are set to a default value,
chosen by the programmers during implementation. The \emph{race-to-idle}
strategy is applied by default on devices, so all the computing resource will be allocated with their highest operation condition. We use the term \setting{DEFAULT} to refer to this default application and device knob setting.

The most common performance metric of a streaming application is the rate of
processing inputs, or \emph{FPS}.
We demonstrate how \setting{DEFAULT} performs by inferencing various deep neural network models on a dual-socket desktop workstation with 64 cores. All the models are implemented within \emph{TensorFlow}~\cite{tensorflow}.
For simplicity, all the application knobs are fixed and only the number of cores to which an application is deployed is controlled in this experiment.

Since \setting{DEFAULT} may not yield the best performance,
we use term \setting{ORACLE} to refer to the optimal knob setting given by exhaustive profiling.
The comparison between \setting{DEFAULT} and \setting{ORACLE} as well as the optimal number of cores to use are shown in Table~\ref{tb:def_bad}.
While \setting{DEFAULT} should yield the best performance in many cases,
surprisingly, no model has the best performance when uses all the 64 cores in this experiment.
There is a geometric mean performance loss of 40\% when using \setting{DEFAULT}, and different models have different \setting{ORACLE} settings.
The primary cause is that the communication overhead between cores grows with the number of cores being used. The \setting{ORACLE} setting achieves better balance between communication and the amount of work allocated to each core.
Also worth to notice that using \setting{DEFAULT} results in unnecessary recourse occupation.
Though \setting{DEFAULT} is easy to use,
this experiment shows that \setting{DEFAULT} may underutilize modern computer systems that have more and more non-trivial parallelism.

\begin{table}[t]
  \centering
  \begin{small}
  \begin{tabular}{l|r|r|r|r}
        Applications & \textbf{DEFAULT} & \textbf{ORACLE} & \textbf{Cores} & \textbf{Speedup}\\
        \toprule
        \textbf{ResNet8}                  & 1409.01 & 1769.18 & 4  & 1.25x \\
        \textbf{ResNet50}                 & 53.46   & 60.88   & 46 & 1.14x \\
        \textbf{MobileNet\_V2}            & 124.57  & 139.02  & 15 & 1.12x \\
        \textbf{Visual wake words}        & 245.11  & 267.25  & 4  & 1.09x \\
        \textbf{Speech recognition}       & 2.06    & 4.26    & 2  & 2.07x \\
        \textbf{Text classification}      & 124.92  & 257.85  & 7  & 2.06x \\
  \end{tabular}
  \end{small}
  \caption{Model inference performance comparison between \setting{DEFAULT} and \setting{ORACLE} on a dual-socket 64-core desktop workstation with TensorFlow.\label{tb:def_bad}}
  \vspace{-20pt}
\end{table}

This experiment involves optimizing different applications on one device without imposing constraints. It becomes more difficult to find the optimal knob setting when having more dimensions of complexity: devices, inputs, user-specified performance objective and constraints (Section~\ref{sec:bg:dev} - Section~\ref{sec:bg:obj_cons}).

\subsection{Device Diversity}
\label{sec:bg:dev}

We demonstrate device diversity by running the same application on two popular embedded platforms:
i) \odroid, ii) \jetson. Both devices are heterogeneous platforms with two types of cores.
The application used is \app{Vips}, a parallel image transformer for large uncompressed images.

\begin{figure}[t]
    \vspace{-15pt}
    \centering
    \subfloat[\odroid]{\includegraphics[scale=0.55]{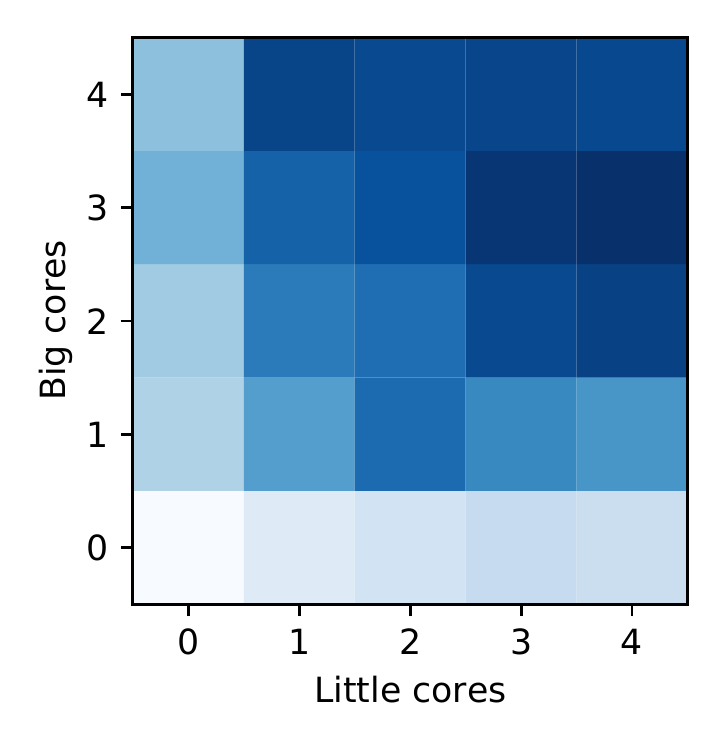}\label{fig:odroid_colorbar}}
    \subfloat[\jetson]{\includegraphics[scale=0.55]{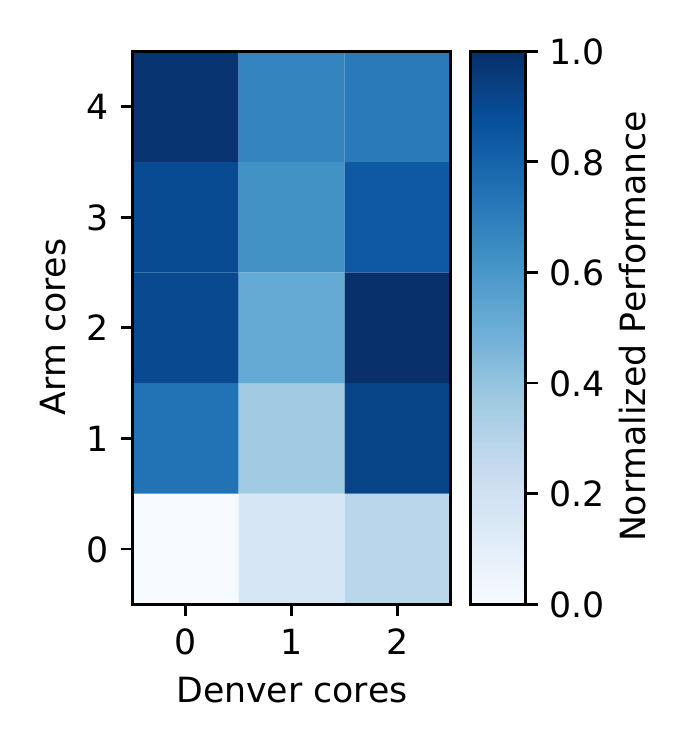}\label{fig:jetson_colorbar}}
    \caption{Performance of \app{Vips} on \odroid and \jetson with different combination of cores.}
    \label{fig:surface}
\end{figure}

Figure~\ref{fig:surface} shows the normalized performance of \app{Vips} on two platforms when using different core combinations.
\app{Vips}'s performance is neither linear nor convex, and has very different patterns between two boards.
First of all, neither boards yield the best performance when using \setting{DEFAULT}.
Second, the optimal knob setting is three big cores and four little cores on
\odroid while two Denver cores and two Arm cores on \jetson. It is not
trivial to obtain the optimal knob setting due to the complexity in application's implementation (e.g. load balancing, data exchange \& sharing) and device's architecture (e.g. compute capacity of/between cores).


\subsection{Input Sensitivity}
\label{sec:bg:inp_sen}
\begin{figure}[t]
    \centering
    \subfloat[Rendered content]{\includegraphics[scale=0.09]{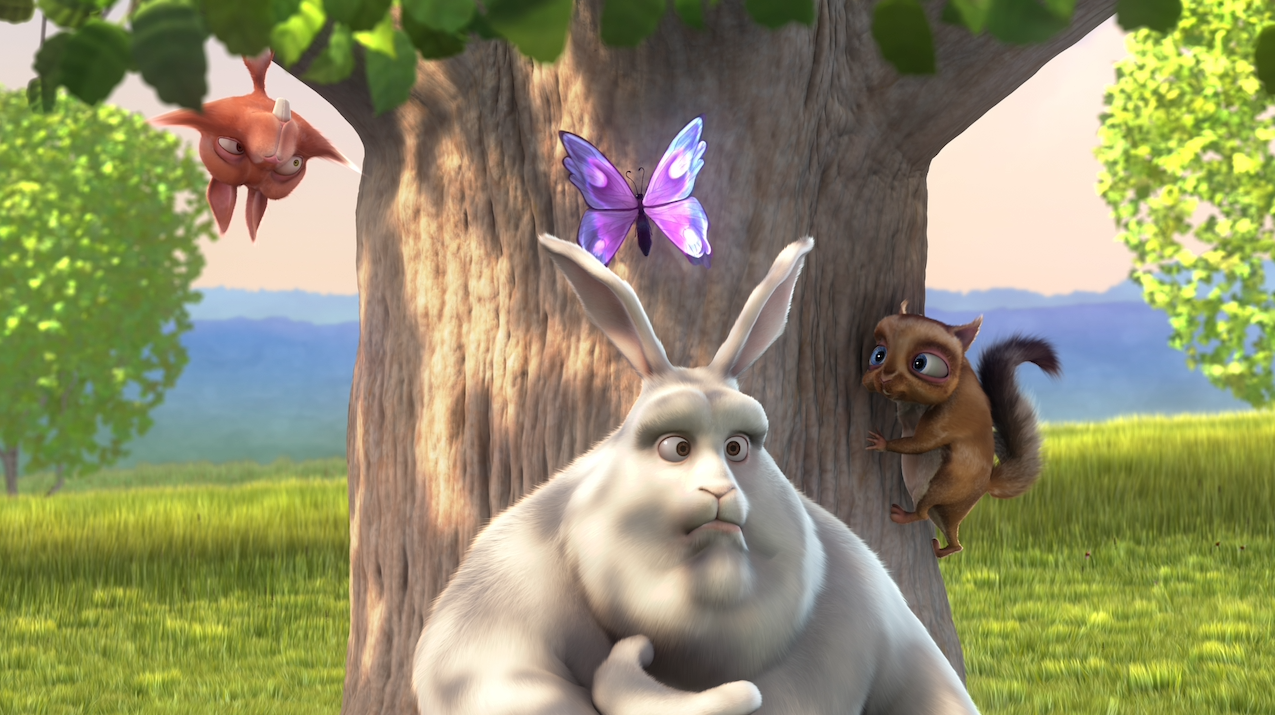}\label{fig:bbb}}
    \subfloat[Photographic content]{\includegraphics[scale=0.09]{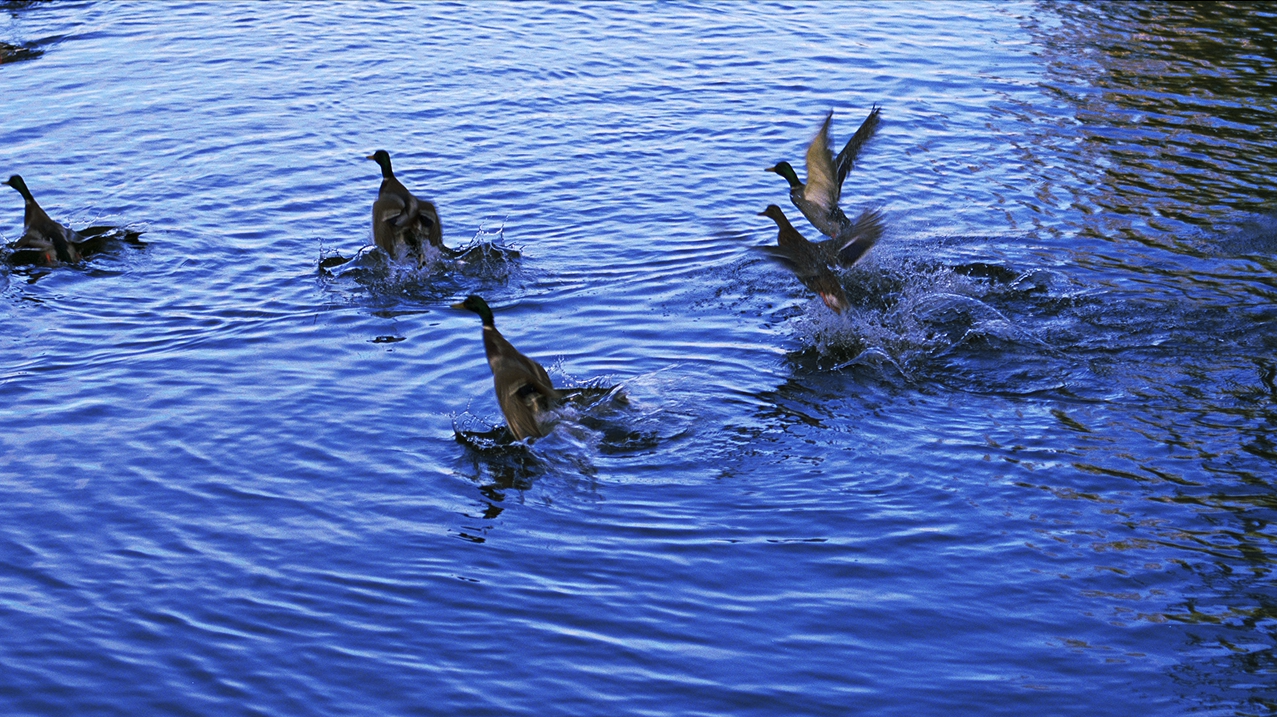}\label{fig:ducks}}
    \caption{Example of content difference of \app{X264} inputs.}
    \label{fig:x264_ex}
\end{figure}

The next layer of complexity comes from inputs. The most straightforward factor
is input size. Video encoder \app{X264} encodes 720p videos $\sim$4$\times$ faster than
1080p videos. More unnoticeable factors involve input content. As shown in
Figure~\ref{fig:x264_ex}, though both videos are of 720p, encoding rendered
content (14.3 FPS) is almost 2$\times$ faster than encoding photographic content (8.3
FPS) on an \odroid board. The potential reason is that rendered content is
relatively ‘easier' because it is generated with algorithms.
Input sensitivity of such leads to distinct optimal knob settings
when having a fixed performance requirement, e.g. minimum FPS requirement.

\subsection{Objectives and Constraints}
\label{sec:bg:obj_cons}

\begin{figure}[t]
    \vspace{-10pt}
    \subfloat[FPS Vs. Power]{\includegraphics[scale=0.53]{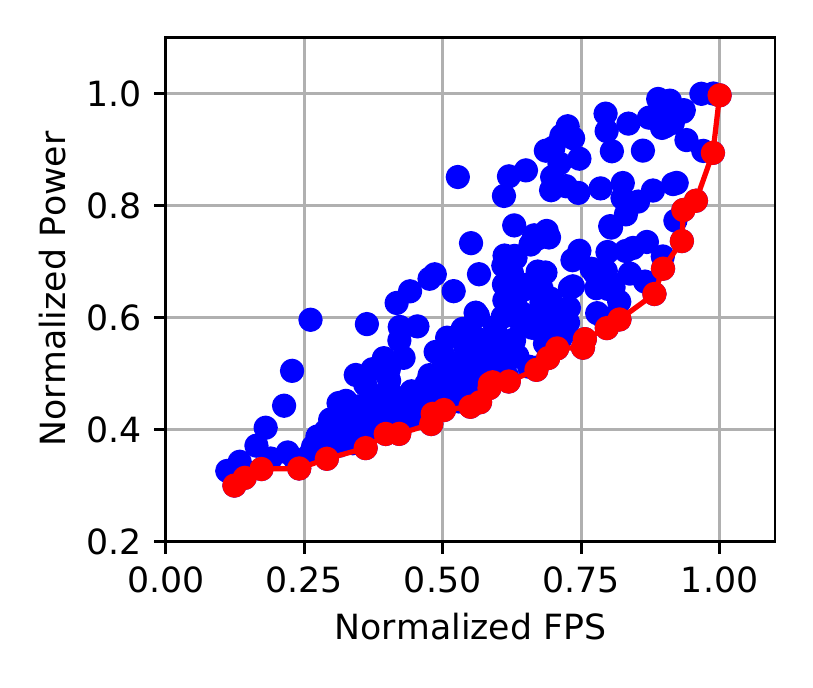}\label{fig:x264_pareto_fps_power}}
    \subfloat[FPS Vs. Energy]{\includegraphics[scale=0.53]{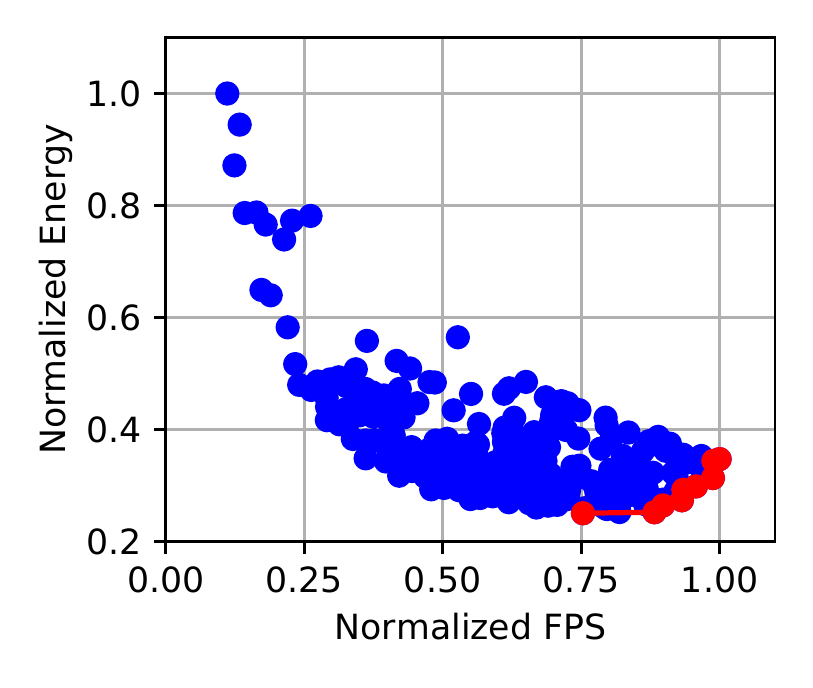}\label{fig:x264_pareto_fps_energy}}
    \caption{Example pareto frontiers of \app{X264} on Odroid XU4. Each point represents one knob setting.}
    \label{fig:pareto}
\end{figure}

User-specified run-time objective and constraint also affect the optimal knob setting.
An objective is defined as a metric to optimize; a constraint is represented by a metric and a set point. Meeting a constraint means the metric value is lower ( or higher) than the set point per user request. Note that constraint is optional and there can be more than one constraints.

In Figure~\ref{fig:pareto}, we display \emph{pareto frontiers} of two
constrained optimization problems of \app{X264} running on \odroid. It is clear that given different constraint metrics, their pareto frontiers are significantly unalike;  given one constraint such as power, different set points result in distinct knob settings on the pareto frontier.
It also worth to notice that there are plenty of ``good enough'' knob settings near the pareto frontier that yield similar performance.

Most prior works~\cite{capri, caloree} focus on solving only one type of objective and constraint combination,
such as ``Optimize performance given an accuracy bound'' or ``Least energy under a performance cap''.
However, objective and constraint can be any meaningful combinations depending on actual
use cases. Other common (constrained) optimization problems includes
``\emph{Maximize FPS}'', ``\emph{Maximize FPS under a power/energy cap}'',
''\emph{Minimize resource usage given an FPS requirement}'', \emph{etc}.



\subsection{Summary}
\label{sec:bg:summary}
In addition to the factors mentioned above, there can be other factors that have non-trivial run-time impact, such as extreme core temperature (self-protection mechanism), running multiple streaming applications. Those situations make the run-time behavior even more unpredictable.

In this section, we show that:
\begin{closeitemize}
    \item In optimization problems (without constraint), \setting{DEFAULT} may deliver sub-optimal performance and cause unnecessary resource occupation. 
    \item The cross product of all the factors makes it difficult if not impossible to derive closed-form analytical expressions for various metrics. Approaches that rely on offline profiling are not feasible (overhead, scope, portability) for solving such control problem.
\end{closeitemize}

The goal of this paper is to explore the idea of online control and develop a
generic methodology to control streaming applications at run time in a
principled fashion so that the objective is optimized while the constraint is
also met.

\section{Problem Formulation}
\label{sec:def}

In this section, we describe the formulation of the constrained optimization problem we target to solve.
We target general streaming applications but will focus on those long-running ones. Applications might have distinct phases with different optimal knob solutions, but the run-time behavior for successive inputs within one phase is relatively stationary. We assume that each phase is long enough ($>$ 1 minute) so that the knob setting provided by the sampling-based online controller can be beneficial.

To keep the notation simple, since a minimization/maximization problem can be easily converted to its opposite problem by multiplying a ``-1'', we assume that the objective metric is maximized and the constraint metric is kept under a set point for demonstration purpose in this paper. We also assume that only one constraint is used, and this can be easily extended to using no constraint or more than one constraints.


Informally, the target control problem can be formulated as the following: for each experiment, a user specifies a \emph{run-time configuration}, including an application, its input(s), a device, a objective and a constraint. Within each phase of the application's execution, find the knob setting at run time in the combined knob space of application and device, so that the objective is maximized and the constraint is met.
This can be formulated as the following constrained optimization problem.


\newtheorem{problem}{\textbf{Problem Formulation}}
\begin{problem}
Given a run-time configuration,
\begin{itemize}
    \item A streaming application $A$ with knob space $\kappa_{A}$
    \item A device $D$ with knob space $\kappa_{D}$
    \item An objective metric $f_o$
    \item An constraint metric $f_c$ and set point $\epsilon$
    \item An input $I$
\end{itemize}

Within each phase of the execution, find $k \in \kappa_A \times \kappa_D$ such that
\begin{itemize}
    \item $o = f_o(A, D, I, k)$ is maximized
    \item $c = f_c(A, D, I, k) < \epsilon$
\end{itemize}
\end{problem}

\section{Sampling-based Online Controller}
\label{sec:design}


\subsection{Choice of Online Control Strategy}
\label{sec:oc_choice}

\subsubsection{\textbf{Traditional Online Control}}
A traditional online control system can be represented as a \emph{supervised}
feedback loop where the system monitors certain quantities of interest and
compares them to their reference values. Base on the difference, the controller
tunes knob settings in the direction of reducing the gap.
Cruise control is a typical online control system that adjusts acceleration by
looking at the difference between vehicle's current speed and a reference speed.
Traditional control systems have already been used to adjust knob settings in
computer systems~\cite{green, rumba, mimo, spectr, ma2011scalable, fu2011cache, slambooster2}.

Though traditional methods have the advantage of formal
reasoning~\cite{ctrl_handbook}, a function for updating knob settings based on
the quantity gap is needed. These functions typically rely on strong structural
assumptions of being linear, convex or deterministic. However, as discussed in Section~\ref{sec:bg:dev}, the relationship between knob settings and performance
can be nonlinear or non-convex, with unknown function form.
Plus the diversity in run-time configurations, it is difficult to come up with a general tuning function.
Additionally, traditional online control requires a reference value to react.

\subsubsection{\textbf{Sampling-based Method}}
Sampling is a popular alternative approach when analytic methods aren't options.
Instead of depending on gaps towards reference values, sampling
is an \emph{unsupervised} approach to pick knob settings only based on the
performance patterns of the already sampled points.
During one sampling phase, a sequence of knob settings are picked incrementally and strategically, of which
each knob setting is decided based on the statistics of all previous sampled points.
Unlike traditional control loop,
sampling-based approach does not require to know the behavior pattern of the
whole knob spaces ahead of time in order to pick a knob setting,
so that it can adapt to any run-time configuration.
Therefore in this paper, sampling approach is chosen as our basic online control strategy.

\subsection{Controller Overview}
\label{sec:infra}

The control flow of \name is presented Algorithm~\ref{alg:controller}. The main
idea of \name is to utilize the beginning portion of each phase for sampling, aiming to find the optimal knob setting for the remaining of that phase.
An experiment starts with initializing the target device and application (line~\ref{alg:init_d} - \ref{alg:init_a}). Both the application and the device report run-time statistics regarding the user-specified objective and constraint.

\begin{algorithm}[t]
    \caption{Sampling-based online control loop for application $A$ executing on device $D$, given input $i$, application knob space $\kappa_A$, device knob space $\kappa_D$,
        objective metric $f_o$, constraint metric $f_c$ and set point $\epsilon$.}
    \label{alg:controller}
    \begin{algorithmic}[1]

        \State $D.\mathit{init()}$ \label{alg:init_d}
        \Comment{Initialize device}
        \State $A.\mathit{start(D)}$ \label{alg:init_a}
        \Comment{Start application on device}
        \State
        \State $\mathit{new\_phase}$ \letequal True
        \Comment{Start with a new phase}
        \While {not $A.\mathit{finished()}$}
            \If {$\mathit{new\_phase}$}
                \State $o, c, k_{best}$ \letequal $\mathit{sampler}(\kappa_A \times \kappa_D, f_o, f_c, \epsilon)$ \label{alg:sampler_start}
                \State $\mathit{setKnob(A, D, k_{best})}$ \label{alg:sampler_end}
                \State $\mathit{new\_phase}$ \letequal False
            \EndIf
                
            \State
            \State $o', c'$ \letequal $\mathit{monitor}()$ \label{alg:pd_start}
            \If {$\mathit{distance}(o, o', c, c') > \delta$}
                \State $\mathit{new\_phase}$ \letequal True \label{alg:pd_end}
            \EndIf
        \EndWhile
	\end{algorithmic}
\end{algorithm}

The main control loop consists of two components: 1) a \emph{sampler}, 2) a \emph{phase detector}.
For each new phase, the sampler samples knob settings from the Cartesian product
of the application and the device knob space. Each sampled
knob setting is evaluated against the user-specified objective and constraint metrics.
At the end of one sampling phase, the knob setting that maximizes the objective
and meets the constraint is chosen and set for the following execution (line~\ref{alg:sampler_start} - \ref{alg:sampler_end}).
The phase detector determines whether a new sampling phase is needed by checking the difference between the current run-time performance with the reference performance of the picked knob recorded during the sampling phase (line~\ref{alg:pd_start} - line~\ref{alg:pd_end}).
The design of the sampler and the phase detector are described in Section~\ref{sec:sampler} and Section~\ref{sec:phase_detect}.

\subsection{Sampler}
\label{sec:sampler}

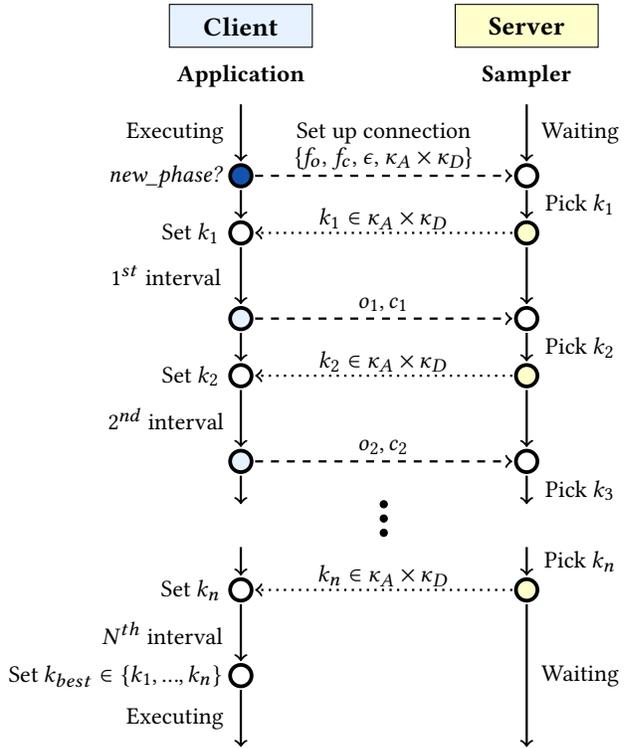
\begin{figure}[t]
  \hspace{-30pt}
  \centering
    \begin{tikzpicture}[scale=0.95]

      \draw [->, line width=0.3mm] (1.5, -0.2) -- (1.5, -1.0);
      \node at (0.58, -0.6) {Executing};

      \draw [line width=0.5mm] (1.5, 0) circle (0.15);
      \node at (-0.25, 0.0) {Set $k_{best} \in \{k_1,...,k_n\}$};

      \draw [->, line width=0.3mm] (5.5, 1.0) -- (5.5, -1.0);
      \draw [->, line width=0.3mm] (1.5, 1.0) -- (1.5, 0.2);
      \node at (0.4, 0.6) {$N^{th}$ interval};
      \node at (6.25, 0.0) {Waiting};

      \draw [line width=0.5mm] (1.5, 1.2) circle (0.15);
      \draw [fill=lightyellow,line width=0.5mm] (5.5, 1.2) circle (0.15);
      \draw [->, dotted, line width=0.3mm] (5.3, 1.2) -- (1.7, 1.2);
      \node at (3.5, 1.4) {$k_n \in \kappa_A \times \kappa_D$};
      \node at (0.8, 1.2) {Set $k_n$};

      \draw [->, line width=0.3mm] (1.5, 1.8) -- (1.5, 1.4);
      \draw [->, line width=0.3mm] (5.5, 1.8) -- (5.5, 1.4);
      \node at (6.25, 1.6) {Pick $k_n$};

      \draw [fill=black] (3.5, 2.0) circle (0.05);
      \draw [fill=black] (3.5, 2.2) circle (0.05);
      \draw [fill=black] (3.5, 2.4) circle (0.05);

      \draw [->, line width=0.3mm] (1.5, 2.8) -- (1.5, 2.4);
      \draw [->, line width=0.3mm] (5.5, 2.8) -- (5.5, 2.4);
      \node at (6.25, 2.6) {Pick $k_3$};

      \draw [fill=lightblue, line width=0.5mm] (1.5, 3.0) circle (0.15);
      \draw [line width=0.5mm] (5.5, 3.0) circle (0.15);
      \draw [->, dashed, line width=0.3mm] (1.7, 3.0) -- (5.3, 3.0);
      \node at (3.5, 3.2) {$o_2, c_2$};

      \draw [->, line width=0.3mm] (1.5, 4.0) -- (1.5, 3.2);
      \draw [->, line width=0.3mm] (5.5, 4.0) -- (5.5, 3.2);
      \node at (0.45, 3.6) {$2^{nd}$ interval};

      \draw [line width=0.5mm] (1.5, 4.2) circle (0.15);
      \draw [fill=lightyellow, line width=0.5mm] (5.5, 4.2) circle (0.15);
      \draw [->, dotted, line width=0.3mm] (5.3, 4.2) -- (1.7, 4.2);
      \node at (3.5, 4.4) {$k_2 \in \kappa_A \times \kappa_D$};
      \node at (0.8, 4.2) {Set $k_2$};

      \draw [->, line width=0.3mm] (1.5, 4.8) -- (1.5, 4.4);
      \draw [->, line width=0.3mm] (5.5, 4.8) -- (5.5, 4.4);
      \node at (6.25, 4.6) {Pick $k_2$};

      \draw [fill=lightblue, line width=0.5mm] (1.5, 5.0) circle (0.15);
      \draw [line width=0.5mm] (5.5, 5.0) circle (0.15);
      \draw [->, dashed, line width=0.3mm] (1.7, 5.0) -- (5.3, 5.0);
      \node at (3.5, 5.2) {$o_1, c_1$};

      \draw [->, line width=0.3mm] (1.5, 6.0) -- (1.5, 5.2);
      \draw [->, line width=0.3mm] (5.5, 6.0) -- (5.5, 5.2);
      \node at (0.45, 5.6) {$1^{st}$ interval};

      \draw [line width=0.5mm] (1.5, 6.2) circle (0.15);
      \draw [fill=lightyellow, line width=0.5mm] (5.5, 6.2) circle (0.15);
      \draw [->, dotted, line width=0.3mm] (5.3, 6.2) -- (1.7, 6.2);
      \node at (3.5, 6.4) {$k_1 \in \kappa_A \times \kappa_D$};
      \node at (0.8, 6.2) {Set $k_1$};

      \draw [->, line width=0.3mm] (1.5, 6.8) -- (1.5, 6.4);
      \draw [->, line width=0.3mm] (5.5, 6.8) -- (5.5, 6.4);
      \node at (6.25, 6.6) {Pick $k_1$};

      \draw [fill=bestblue, line width=0.5mm] (1.5, 7.0) circle (0.15);
      \draw [line width=0.5mm] (5.5, 7.0) circle (0.15);
      \draw [->, dashed, line width=0.3mm] (1.7, 7.0) -- (5.3, 7.0);
      \node at (3.5, 7.25) {\{$f_o$, $f_c$, $\epsilon$, $\kappa_{A} \times \kappa_{D}$\}};
      \node at (3.5, 7.6) {Set up connection};
      \node at (0.45, 7.0) {\textit{new\_phase?}};

      \draw [->, line width=0.3mm] (1.5, 8.0) -- (1.5, 7.2);
      \draw [->, line width=0.3mm] (5.5, 8.0) -- (5.5, 7.2);
      \node at (0.58, 7.6) {Executing};
      \node at (6.25, 7.6) {Waiting};


      \node at (1.5, 8.4) {\textbf{Application}};
      \node at (5.5, 8.4) {\textbf{Sampler}};

      \draw [fill=lightblue] (0.5, 8.8) rectangle (2.5, 9.4);
      \draw [fill=lightyellow] (4.5, 8.8) rectangle (6.5, 9.4);

      \node at (1.5, 9.1) {\textbf{\large Client}};
      \node at (5.5, 9.1) {\textbf{\large Server}};

    \end{tikzpicture}
    \caption{Sampling process}
    \label{fig:sample_flow}
\end{figure}

The flow of one sampling phase with $N$ rounds is shown in Figure~\ref{fig:sample_flow}.
In order not to add disturbance to application's run-time behavior, the target application and the sampler are deployed separately onto a client (target device) and a server. The client and the server communicate by network connections.

A new sampling phase is activated when a new experiment begins or the phase detector identifies a new phase. Upon start, the client sets up a connection with the server by sending the objective ($f_o$), the constraint ($f_c, \epsilon$) and the knob space ($\kappa_A \times \kappa_D$) of the current run-time configuration to the server.
The first round of the sampling phase starts by picking the first knob ($k_1$) on the server and sending it back to the client.
Upon receiving $k_1$, the client sets $k_1$ for the application and the device,
and measures its run-time statistics during one measurement interval ($\sim$3 seconds in our experiments). After measuring $k_1$, its objective and constraint metric values $o_1$ and $c_1$ are sent back to the server. The streaming application keeps running with $k_1$ before it hears back from the server again. The server then calculates the second knob setting and repeats the above procedure.
After the $N^{th}$ knob setting is measured, the knob setting that maximizes the objective and meets the constraint will be picked and set for the remaining execution until a new execution phase is detected.

The strategy to sample the knob space is the key to find a competitive knob
setting. More number of samples generally means
better knob setting selection, but evaluating samples take time, and cause
negative impact when evaluating sub-optimal samples.
More samples also lead to less time for the chosen knob setting to be beneficial
because applications have finite execution duration.
In this paper, only limited number of
samples is allowed to be drawn ($N \ll |\kappa_A \times \kappa_D|$).
$N$ is chosen to be between 8 and 12 given
application's overall execution duration and knob space size.

In order to get a good grasp of the target knob space with small number of
samples, we use \emph{global optimization} techniques to design the sampling process.
The key is to utilize the knowledge given by all previous samples,
while balancing the dilemma of \emph{exploration} and \emph{exploitation}.
Exploration generally means probing under-sampled portions of the knob space, for
the purpose of getting out of local optimums and finding promising regions are yet to be sampled.
Exploitation, on the other hand, searches a known promising region
hoping to find a better sample than the samples already drawn
from this region. By leveraging exploitation and exploration trade-off, the sampler can
quickly identify unpromising regions and focus on promising regions only.
Prior works usually attempt to build the performance model of the whole
knob space, which is neither necessary nor cheap~\cite{hbm, caloree}.
Locating promising regions quickly is also beneficial to at least get a near
optimal solution because there can be plenty of knob settings near the pareto
frontier that yield similar performance (\emph{e.g.} Figure~\ref{fig:pareto}).

Following the spirit of balancing exploration and exploitation, the whole sampling
phase of \name is divided into two stages: 1) \emph{initialization}; 2) \emph{searching}.

\subsubsection{\textbf{Initialization}}
\label{sec:init}
The initialization stage consists of the first $M$ of the total $N$ samples.
In this stage, \name aims to learn the knob space at a coarse scale,
so all sample choices follow the principle of exploration.
Too few samples may cause exploitation in wrong regions while too many
initial samples may leads to insufficient exploitation during the searching stage.
Common choices of sampling strategy for the initialization stage includes \emph{random} sampling and \emph{Latin hypercube} sampling (LHS)~\cite{lhs}.
Taking a 2$D$ knob space as example, each sample by LHS marks in which row and
column the sample was taken and subsequent samples avoids those marked rows and columns.
In contrast to \naive random sampling, LHS ensures that the set of picked knob settings is representative of the real variability.

\subsubsection{\textbf{Searching}}
\label{sec:search}
\begin{figure}[h]
  \centering
    \begin{tikzpicture}[scale=1.0]
      \draw [-, line width=0.4mm] (0.0, 1.3) -- (8.0, 1.3);

      \draw [-, line width=0.4mm] (0.0, 1.3) -- (0.0, 1.45);
      \draw [-, line width=0.4mm] (2.5, 1.3) -- (2.5, 1.45);
      \draw [-, line width=0.4mm] (8.0, 1.3) -- (8.0, 1.45);

      \node at (0.0, 1.0) {$0$};
      \node at (2.5, 1.0) {$M$};
      \node at (8.0, 1.0) {$N$};

      \node at (1.25, 1.6) {\textbf{Initialization}};
      \node at (5.25, 1.6) {\textbf{Searching}};

    \end{tikzpicture}
    \caption{Sampling stages}
    \label{fig:init_search}
\end{figure}
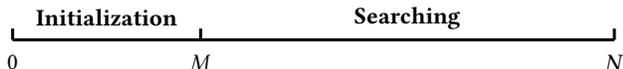

The remaining $M-N$ samples all belongs to the
searching stage.
The goal of this stage is to find the knob setting that
maximizes the objective while meeting the constraint through exploration and exploitation.
The searching stage can be implemented in multiple ways, discussed in the following section.
The arrangement of initialization and searching in a sampling phase is
shown in Figure~\ref{fig:init_search}.

\subsection{Searching Strategies}
\label{sec:op_strat}

\subsubsection{\textbf{Random sampling \& LHS}}
Random sampling and LHS can also be used in the searching stage.
However, neither of them utilizes the history of sampled knob settings since
their samples can be generated given only random seed and knob space.
These two sampling strategies are not helpful to find the optimal knob setting
because no exploitation but only exploration is involved.

\subsubsection{\textbf{Regressors}}
\label{sec:reg_intro}
The second searching strategy is built on machine learning regressors (models).
Popular types of regressors includes \emph{linear regressors}~\cite{linear_reg},
\emph{ensemble regressors}~\cite{random_forest, gradient_boost},
and \emph{Gaussian process regressors}~\cite{gp_reg}, \emph{etc}.
In a sampling phase, a regressor is initialized on the data of all the sampled knob
settings of the initialization stage and is updated after each newly sampled knob setting gets evaluated.

For each next knob setting in the searching stage, it is decided by first
predicting every unsampled knob setting's objective and constraint value
with the most recently updated regressor, and then pick the one that maximizes the objective metric while
staying under the constraint. After this sampled setting get evaluated,
the regressor will get updated or rebuilt based on its type.
Since the number of samples is small, rebuilding the regressor
does not introduce noticeable overhead on a server.
This process repeats for all $N - M$ samples.

In contrast to random sampling and LHS, regressors does utilize history information.
However, this strategy only has exploitation because it takes
the sample that maxes out the prediction at every round.
When a new sample's performance aligns well with the regressor's
prediction, this new sample won't provide new information to update the regressor,
causing the following sample to be very close to this one.
This leads to difficulty in getting out of local minimums.
Extra exploration is needed for better knob choices in the searching stage.

\subsubsection{\textbf{Bayesian optimization (with unknown constraint)}}
$\newline$
Another searching strategy that fits our control problem well is
\bo~\cite{bo, bo_tutorial, bo_review}.
\bo is a sequential design strategy for derivative-free global
optimization of black-box functions that are expensive or lengthy to evaluate
(\emph{e.g.} hyperparameter tuning for machine learning models~\cite{feurer2019auto, snoek2012practical, snoek2015scalable}).


Abstractly, \bo aims to solve the problem of maximizing a target function $f: \mathbb {R} ^{n}\to \mathbb {R}$ (\emph{i.e.} objective metric $f_o$) over a feasible set $X$ (\emph{i.e.} knob space $\kappa_A \times \kappa_D$).
\bo consists of two main components: a probabilistic surrogate model –most commonly, a Gaussian process regression (GP) –that
models the objective function, and an acquisition function (\emph{e.g. Expected
Improvement}~\cite{bo_tutorial}) for deciding where to sample next.

After the initialization phase, \bo construct a GP model based the initial samples.
Given a point $x$ in $X$, GP models $f(x)$ as a normal distribution with
mean $\mu_x$ and variance $\sigma^2_x$.
The more samples taken in one region,
the smaller variance the points of this region tends to have.
When determining the next sample, the acquisition function takes both the mean
and the variance into consideration, in order to tradeoff exploration (large variance) and exploitation
(large mean). The GP model is incrementally updated as new samples are evaluated.


Constraints can be incorporated into \bo by scaling the acquisition function at
each $x$ with its probability of meeting the constraint~\cite{bo_unknown}.
For each constraint, a separate GP model is built. The estimated probability of
one point $x$ to meet a certain constraint, $P(f_c(x) < \epsilon)$, is done by
evaluating the \emph{cumulative distribution function} at $x$
on that GP model.



\subsubsection{\textbf{Hybrid approach}}
\label{sec:hybrid}
Though \bo fits the control challenge very well,
our experiments shows that it often requires a considerable number of samples to
converge to a good solution. For example, \bo tends to focus too much on exploration when
taking 12 samples in a knob space that consists of more than 1200 knob settings.

In order to encourage exploitation in \bo,
\name leverages a hybrid strategy of combining regressors into \bo to enforce
exploitation during the searching stage. We choose GP regressor because
\bo's underlying model is also GP.

\begin{figure}[h]
  \centering
    \begin{tikzpicture}[scale=1.0]
      \draw [-, line width=0.4mm] (0.0, 1.3) -- (8.0, 1.3);

      \draw [-, line width=0.4mm] (0.0, 1.3) -- (0.0, 1.45);
      \draw [-, line width=0.4mm] (2.0, 1.3) -- (2.0, 1.45);
      \draw [-, line width=0.4mm] (2.5, 1.3) -- (2.5, 1.45);
      \draw [-, line width=0.4mm] (7.5, 1.3) -- (7.5, 1.45);
      \draw [-, line width=0.4mm] (8.0, 1.3) -- (8.0, 1.45);

      \node at (0.0, 1.0) {$0$};
      \node at (2.0, 1.0) {$M$};
      \node at (8.0, 1.0) {$N$};

      \node at (1, 1.6) {\textbf{LHS}};
      \node at (2.25, 1.6) {\textbf{GP}};
      \node at (5, 1.6) {\textbf{Bayesian optimization}};
      \node at (7.75, 1.6) {\textbf{GP}};

    \end{tikzpicture}
    \caption{Hybrid approach}
    \label{fig:hybrid}
\end{figure}
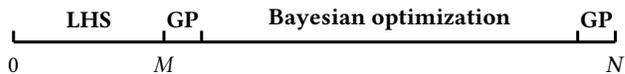

The design of the hybrid approach is shown in Figure~\ref{fig:hybrid}. The
sampler starts with Latin hypercube sampling (LHS) to perform the initial
exploration of the knob space.
At the beginning of the searching stage, a GP regressor is built on the initial
samples and one next sample is generated using this GP regressor.
This sample can potentially boost the sampling quality of \bo because
it provides an ``okay'' solution so that unpromising regions are much easier to
identify, leading to less exploration needed.
For the last sample, \bo may still bias to exploration based on its acquisition
function but exploration is not beneficial at the last step.
In \name, we enforce exploitation by building a GP regressor based on all the
previous samples to generate the final knob setting sample.


\subsubsection{\textbf{Discussion}}
\label{sec:choice_go}
The primary reason to choose \bo over other global optimization techniques is that
it construct a surrogate model incrementally with very limited number of samples.
Instead of assuming strong structural assumptions of being linear and convex, 
\bo assumes that closer
knob settings should have more similar performance by defining a covariance function
(\emph{e.g. RBF kernel}\footnote{https://en.wikipedia.org/wiki/Radial\_basis\_function},
\emph{Mat\'{e}rn kernel}\footnote{https://en.wikipedia.org/wiki/Mat\'{e}rn\_covariance\_function}).
Global optimization techniques such as
simulated annealing~\cite{simulated_annealing}, evolutionary
algorithm~\cite{evolutionary} assume no functional structure at all.
However, those techniques require much more samples to converge to a good
solution, which is not feasible for the control problem in this paper.


In the domain of reinforcement learning,
upper confidence bound (UCB) and Thompson sampling~\cite{thompson} are
two popular heuristics to deal with the ``exploration
Vs. exploitation'' dilemma for problems like the multi-armed bandit
\footnote{https://en.wikipedia.org/wiki/Multi-armed\_bandit}.
However, those techniques usually require multiple visits to every knob setting
in order to be effective, which far exceeds the number of samples allowed in
this paper.

\subsection{Phase Detection}
\label{sec:phase_detect}

After the sampling phase, the best knob setting among all the sampled ones
according to the user-specified objective and constraint is set for the application and the device.
In order to detect new phase, the client keeps reporting run-time statistics to the server.
For each measuring interval, the phase detector compares the received run-time
performance with the recorded performance of the chosen knob setting during
the sampling phase. If the different is larger than 10\% and lasts for two
consecutive intervals, a new sampling phase will be activated to search for a
new solution.



\subsection{Implementation Details}
\label{sec:imple}

\paragraph{\textbf{Standalone implementation}}
\name is designed and implemented independent of any application, device, input, objective and constraints.
Other than reporting performance at run time, no change is needed for
the target application and device.

\paragraph{\textbf{Avoid duplicated samples}}
While it is rare to see duplicated samples in continuous knob space, it's common
in discrete knob space because knob values are rounded to the nearest discrete value.
Duplicated samples waste precious attempts of drawing samples.
A nearby knob setting that has the highest acquisition value will be selected
when duplication happens.

\paragraph{\textbf{Avoid system instability}}
In order to avoid the instability caused by turning on/off cores,
we change application's core allocation only by setting its thread affinity.
Since changing thread affinity has delays and introduce extra noise into measurements,
knob settings are ordered so that the total distance between successive knob settings are
minimized (i.e. gray code encoding) during the initialization stage.
Furthermore, we also make the default knob setting be the first sample point.
In this way, it is possible to tackle minimization problems with requirements represented by ratios.


%
%
%
%
%

\section{Experimental Evaluation}
\label{sec:eval}

\subsection{Experiment Setup}
\label{sec:setup}

\subsubsection{\textbf{Benchmarks and Inputs.}}
\label{sec:bench}

In this work, \name is evaluated on two benchmark suites. The first one is the
PARSEC benchmark suite~\cite{parsec}, an open-source parallel benchmark suite
for emerging applications for evaluating parallel systems. Benchmarks in PARSEC
cover popular application domains including financial, computer vision, physical
modeling, future media, \emph{etc}. All the applications are implemented in C/C++.
Inputs for different applications are included respectively in this benchmark suite.
Here is the list of parallel applications that are regarded as streaming applications:
\begin{closeitemize}
    \item \emph{\textbf{Bodytrack.}} Bodytrack is a computer vision application that tracks a markerless human body. It helps machines interact with environment better with no aid provided. The inputs to Bodytrack are four sequences of frames fed from 4 cameras.
    \item \emph{\textbf{FaceSim.}} FaceSim is a computer graphic application that simulates motions of a human face for visualization purposes. It is used in video games and other interactive animations require visualization of realistic faces in real time. FaceSim takes a series of muscle activation as inputs and simulates on a face model.
    \item \emph{\textbf{FluidAnimate.}} FluidAnimate is a computer animation application that simulates the underlying physics of fluid motion for real-time animation purposes with smoothed particle hydrodynamics algorithm\cite{sph}. It is a highly demanded feature that allows significantly more realistic animations in games and other media productions. The inputs to FluidAnimate is a list of particles.
    \item \emph{\textbf{StreamCluster.}} StreamCluster is a data mining application that computes an approximation for the optimal clustering of a stream of data points. It is a common problem in many fields like network security or pattern recognition. StreamCluster takes a stream of multidimensional points as inputs.
    \item \emph{\textbf{Vips}.} Vips is a image processing that applies a series of transformations to an uncompressed image. Vips primarily targets huge professional images that needs to be handled quickly.
    \item \emph{\textbf{X264}.} X264 is a video encoder that encodes a sequence of images into the H.264/MPEG-4 AVC compression format. It is widely used nowadays because encoded videos are more network bandwidth and storage friendly.
\end{closeitemize}


The second set of benchmarks evaluated is a modified version of MLPerf inference
benchmark suite~\cite{mlperf} that covers the deep learning aspect of streaming applications.
In order to fit and execute on resource-constrained devices, some of the applications are replaced with simpler versions and smaller inputs.
The selected applications include \emph{\textbf{ResNet8}} \& \emph{\textbf{ResNet50}}~\cite{resnet} (image classification),
\emph{\textbf{MobileNet V2}}~\cite{mobilenetv2} (object detection),
\emph{\textbf{visual wake words}} (\emph{MobileNet V1}~\cite{mobilenet}),
\emph{\textbf{speech recognition}} (convolutional neural network) and
\emph{\textbf{text classification}} (recurrent neural network).
All the applications are implemented and inferenced using Tensorflow~\cite{tensorflow}.

Due to the limitation of tuning application knobs due to their availability
(Section~\ref{sec:combined_space}), application knobs are not controlled by default otherwise explicitly mentioned.

\subsubsection{\textbf{Device and knob space.}}
\label{sec:device}
We evaluate \name extensively on multiple platforms with different architectures.

\begin{closeitemize}
    \item \emph{\textbf{\odroid}} is a 32-bit embedded platform with heterogeneous
        big.LITTLE core architecture. \odroid is equipped with four Cortex-A15
        cores and four Cortex-A7 cores with frequency ranging from 0.2 to 2~GHz
        and 0.2 to 1.5~GHz (0.1GHz steps) respectively.
        The power dissipation of the whole board is measured by an external SmartPower2 device.
        The total knob space consists of 6384 knob settings (4 knobs).

    \item \emph{\textbf{\jetson}} is a 64-bit embedded AI computing device that has four
        Cortex-A57 cores and two Denver2 cores that both operate from 0.345
        to 2~GHz (0.15GHz steps).
        The power dissipation of different components of these board is measured through its internal i2C interface.
        Jetson's knob space has 1694 knob settings (4 knobs).

    \item \emph{\textbf{\xeon}} is a modern desktop system with two sockets and 32 physical cores (64 thread cores).
        The power dissipation of this platform is measured by pyRAPL, a software toolkit that uses the "Running Average Power Limit" technology to measure the energy footprint of a host machine along application's execution.
        Only one knob with 64 settings are tuned on this platform.
\end{closeitemize}

%


\subsubsection{\textbf{Metric.}}
\label{sec:metric}
The goal of an online controller is to find the best knob setting possible
to optimize a user-specified objective while staying under constraints if specified.
Due to the randomness and uncertainty involved in the sampling process and measurement,
we evaluate the effectiveness of an online controller by calculating the expectation
of the objective when the constraint is met across independent runs.

$QoS$ (\emph{Quality of service}) is defined as the ratio of the
expected objective between using an online controller and the optimal knob
setting. $QoS$ expressions for maximization and minimization
problems are given in Equation~\ref{eqn:qos_max} and \ref{eqn:qos_min} respectively.
\begin{align}
    QoS_{max} = \frac{E_{ctrl}[o | c < \epsilon]}{E_{op}[o | c < \epsilon]} \times 100\%
    \label{eqn:qos_max}
\end{align}

\begin{align}
    QoS_{min} = \frac{E_{op}[o | c < \epsilon]}{E_{ctrl}[o | c < \epsilon]} \times 100\%
    \label{eqn:qos_min}
\end{align}


\subsubsection{\textbf{Experiment Design.}}
\label{sec:eval:exp_design}
We compare \name with six other control settings. The implementation
of SGD regressor and random forest regressor are from the scikit-learn package
\footnote{https://scikit-learn.org/}. \bo's implementation comes from
package Emukit\footnote{https://github.com/EmuKit/emukit}\cite{emukit}.
\begin{closeitemize}
    \item Race-to-Idle (\setting{DEFAULT})
    \item Random sampling
    \item Linear regressor: SGD
    \item Ensemble regressor: random forest
    \item \bo
    \item \name: \bo + GP regressor
    \item Optimal knob setting (\setting{ORACLE})
\end{closeitemize}

The length of one sampling phase is set to be 12 rounds on \odroid, 10 on
\jetson and 8 on the desktop system according to their knob space size.
Discussions on the impact of the number of samples are discussed in Section~\ref{sec:len_sampling}.
During the sampling phase, due to the fact that command \app{taskset} needs time (0.5$\sim$1s) to stabilize, we set the measurement interval to 3 seconds, in order to balance the length of one sampling interval and the measurement noise.
For fair comparison, the length of the sampling phase is normalized to be 10\% of the total execution time.
$QoS$ is averaged over 40 independent runs for each run-time configuration.


\subsection{QoS on \odroid}
\label{sec:eval:perf_odroid}

\begin{figure*}[t]
    \centering
    \subfloat[$QoS$]{\includegraphics[scale=0.67]{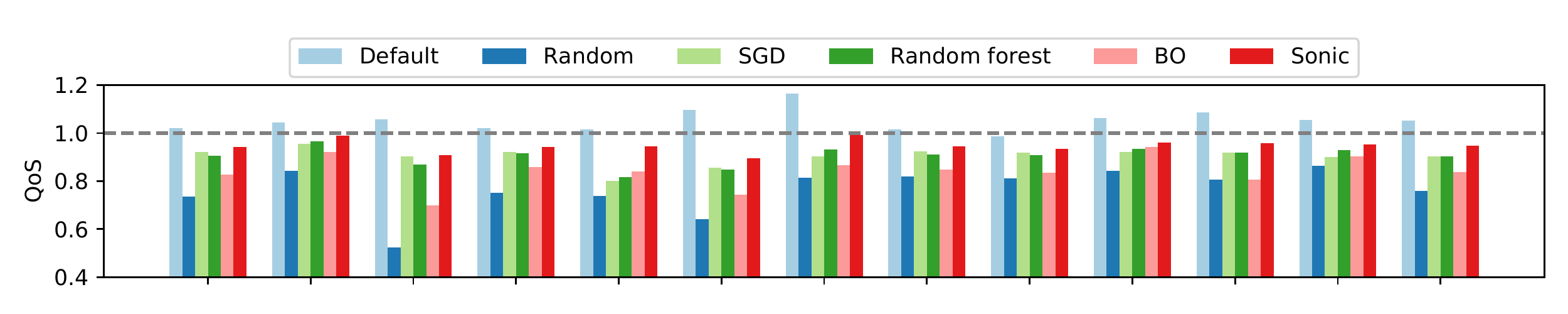}\label{fig:fps_p7_perf}}
    \vspace{0pt}
    \subfloat[Power consumption]{\includegraphics[scale=0.67]{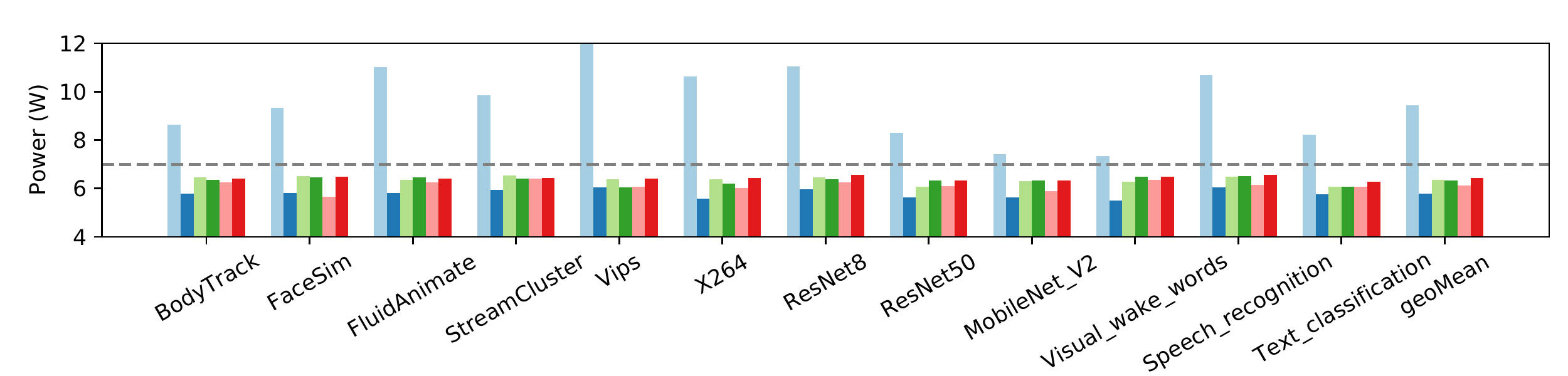}\label{fig:fps_p7_cons}}
    \caption{Application's objective and constraint performance on Odroid XU4.}
    \label{fig:odroid_perf}
\end{figure*}

In this section, we show how \name is able to solve a commonly seen constrained optimization problem: ``Best performance under a power cap of 7~W'' on \odroid for all the applications with their provided inputs in the benchmark suite.

\subsubsection{\textbf{Optimal knob settings.}}
Given this constrained optimization problem, online control would be unnecessary
if all applications' optimal knob settings are the same on one
device. In Table~\ref{tb:best_knob}, we display the optimal knob setting for each
application. For a total of 12 applications, almost every application has a
unique optimal knob setting on \odroid, showing that learning the device alone is
not sufficient in solving such constrained optimization problems.
\begin{table}[!t]
   \centering
      \begin{small}
      \begin{tabular}{l|c|c}
          & \multicolumn{2}{c}{\textbf{Optimal knob setting (ORACLE)}} \\
        \midrule
        \emph{Platform}  & Odroid XU4 & Jetson TX2 \\
        \emph{Objective} & Perf. $\uparrow$ &  Power $\downarrow$ \\
        \emph{Constraint} & Power $<$ 7W & Perf. > 60\% Default \\
      \toprule
        \textbf{Bodytrack}          & (4, 0, 1500, 200)  & (4, 2, 1263, 1263) \\
        \textbf{FaceSim}            & (4, 2, 1400, 1500) & (4, 2, 961,  1263) \\
        \textbf{FluidAnimate}       & (4, 4, 1400, 1500) & (4, 2, 1263, 1263) \\
        \textbf{StreamCluster}      & (4, 4, 1200, 1500) & (4, 2, 961,  1152) \\
        \textbf{Vips}               & (2, 4, 1600, 900)  & (2, 0, 1263, 345)  \\
        \textbf{X264}               & (3, 4, 1100, 1500) & (3, 2, 1263, 1416) \\
      \midrule
        \textbf{ResNet8}            & (4, 4, 1400, 1500) & (3, 2, 1416, 1263) \\
        \textbf{ResNet50}           & (4, 4, 1800, 1500) & (3, 2, 1263, 1263) \\
        \textbf{MobileNet V2}       & (4, 2, 1800, 1500) & (3, 2, 1263, 1263) \\
        \textbf{Visual wake words}  & (4, 2, 1900, 1200) & (3, 0, 1722, 345)  \\
        \textbf{Speech recognition} & (4, 2, 1300, 1500) & (2, 2, 1263, 1263) \\
        \textbf{Text classification}& (3, 0, 1800, 200)  & (4, 0, 1722, 345) \\
      \end{tabular}
      \end{small}
  \vspace{10pt}
  \caption{The optimal knob setting of each application on two platforms given
    different objective and constraint. The first two numbers in the tuple
    represents core numbers, and the last two represents core's frequency.}
  \label{tb:best_knob}
\end{table}

\subsubsection{\textbf{Controller Comparison.}}
The $QoS$ and power dissipation of different controllers are shown
in Figure~\ref{fig:fps_p7_perf} and \ref{fig:fps_p7_cons}.
For \setting{DEFAULT}, all the applications have better performance than \setting{ORACLE}'s.
However, the power dissipation of \setting{DEFAULT} violate the 7
Watt constraint for every application.
Note that the power dissipation of \setting{DEFAULT} is different among
applications, indicating different optimal knob settings given a fixed power
constraint.

When applying sampling-based online controllers, all are able to
provide knob settings that meet the 7-Watt constraint.
The reason is that only knob settings that
meet constraint are considered at the end of each sampling phase.
As expected, random sampling has the worst performance because it does not learn from previous samples.
Regressor-based approach performs much better than random sampling due to their
aggressive exploitation.
SGD regressor does explore regions of interest effectively for many applications,
but not accurate enough due to its underline linear model because the relationship
between performance and knob settings may not be linear.
Random forest regressor is an ensemble regressor that fits a number of decision
trees on various subset of the samples and uses voting to improve the predictive
accuracy and control over-fitting.
Since no assumption on linearity or convexity, it performs better than the SGD regressor.
In terms of the \bo controller, it does trade off exploitation and exploration
during the searching stage, but the knob space is relatively large comparing to
the number of samples, it fails provide enough exploitation to refine the
solutions from exploration.

Finally, after manually adding exploitation at the beginning and the end of
searching stage, \name is able to provide near-optimal results
with only 12 samples in a knob space of 6384. It outperforms regressor-based
approaches because they lack exploration in the searching stage.
Comparing with \setting{ORACLE}, using \name for streaming applications only
incurs $\sim$4.8\% $QoS$ loss. \name achieves this without any prior knowledge of
the run-time configuration. This 4.8\% $QoS$ loss is accounted by two parts:
1) knob settings evaluated during the sampling phase are suboptimal;
2) the knob setting given by the sampler may not be the global optimal.

\subsubsection{\textbf{Individual Experiments.}}
\begin{figure*}[!h]
    \vspace{-10pt}
    \centering
    \includegraphics[scale=0.61]{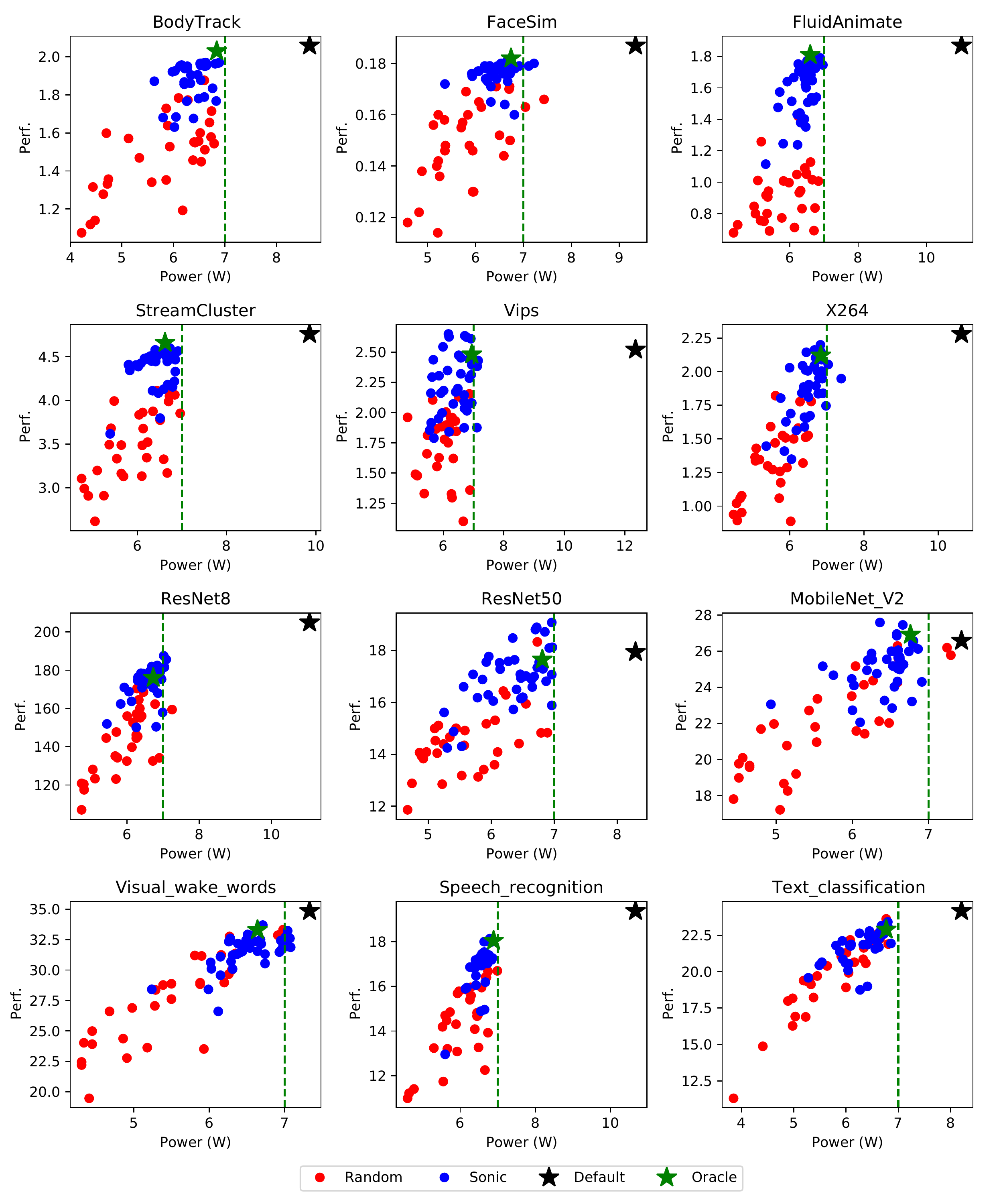}
    \caption{Performance distributions of 40 independent experiments.\label{fig:distributions}}
\end{figure*}

In Figure~\ref{fig:distributions}, we show the objective and constraint performance
of each individual run for all the applications. The average objective and
constraint performance of \setting{DEFAULT} and \setting{ORACLE} is represented by blue and green stars.
Blues dots and red dots represent independent experiments with \name and random sampling respectively.
Using \name significantly improves the objective while still staying under the constraint.
In addition, \name helps reduce the performance variance of individual
runs. Note that uncertainty is involved throughout each experiment due to
randomness and  noise. Though \name can generally boost
performance, it is still possible to yield performance that is worse than other
baseline sampling methods in a single run.

In this experiment, we also observe that \name is not able to find the exact
global optimal knob settings all the time given large knob space, limited number of samples and
uncertainty. However, \name can still reach near-optimal
$QoS$ because there exist plenty of ``good enough'' knob settings
near the pareto frontier that have similar performance (\emph{e.g.} Figure~\ref{fig:pareto}).
In other words, \name is an approximate run-time solution to the optimal knob setting given any run-time configuration.



\subsection{QoS on \jetson}
\label{sec:perf_jetson}
In order to evaluate the ability to solve various constrained optimization problems,
we experiment with a different problem on \jetson:
``Least energy consumption over a performance requirement of 60\%
\setting{DEFAULT}'s performance''.

Due to lack of space, we cannot include figures in this submission.
For this problem, \setting{DEFAULT} can easily meet the performance requirement
but its $QoS$ (minimization problem) of energy is only 0.73. The $QoS$ for
random sampling, SGD regressor, random forest regressor and \bo are 0.81, 0.89,
0.91 and 0.86 respectively.
Finally, \name's $QoS$ of energy is 0.94. The corresponding optimal knob settings of this
problem are included in Table~\ref{tb:best_knob}.

\subsection{QoS on Intel Xeon Gold}
\label{sec:perf_xeon}

In Section~\ref{sec:baseline_perf}, we motivate that \setting{DEFAULT},
which occupies all the cores automatically, are sub-optimal in terms of
performance on a desktop platform for deep learning frameworks.
\setting{DEFAULT} also results in unnecessary compute resource waste.

After applying \name to those applications, a geometric mean of 1.32$\times$ speed up
is achieved ``for free'' because no extra optimization or profiling is applied
to those models. What changed is just the run-time core number selection.
In addition, \name helps reduce resource utilization on this desktop system by 78\% on average,
providing chances for other applications to run on the same device.
When comparing to \setting{ORACLE}, the average $QoS$ loss is 5.5\%.


\begin{table}[t]
  \centering
  \begin{small}
  \begin{tabular}{l|r|r|r|r}
        Applications & DEFAULT & \name & Speedup & $QoS$\\
        \toprule
        ResNet8                  & 1409.01 & 1686.35 & 1.20x & 0.95 \\
        ResNet50                 & 53.46   & 58.64   & 1.10x & 0.96 \\
        MobileNet\_V2            & 124.57  & 134.46  & 1.08x & 0.97 \\
        Visual wake words        & 245.11  & 255.26  & 1.04x & 0.96 \\
        Speech recognition       & 2.06    & 3.73    & 1.81x & 0.87 \\
        Text classification      & 124.92  & 246.95  & 1.98x & 0.96 \\
        Avg. $QoS$ Loss          &  -      & -       & 1.32x & 0.94
  \end{tabular}
  \end{small}
  \vspace{5pt}
  \caption{Model inference FPS comparison between the \setting{DEFAULT}
  and \name on a 2-socket 64-core machine. \label{tb:proteus_ctrl}}
  \vspace{-20pt}
\end{table}

%
%
%
%

\subsection{Phase Detection}
\label{sec:eval:phase}

Some streaming applications may have different phases in one run. It can be another streaming application entering the system, input change and so on. For experimental purpose, we concatenate the ``Big Buck Bunny'' and the ``Ducks take off'' video mentioned in Section~\ref{sec:bg:inp_sen}. They have different performance patterns due to their content difference. The run-time configuration can be described as ``Given certain input, run X264 on \odroid and find the knob setting that optimizes power dissipation while keeping FPS over 2''.

\begin{figure}[t]
    \vspace{-5pt}
    \centering
    \subfloat[FPS]{\includegraphics[scale=0.6]{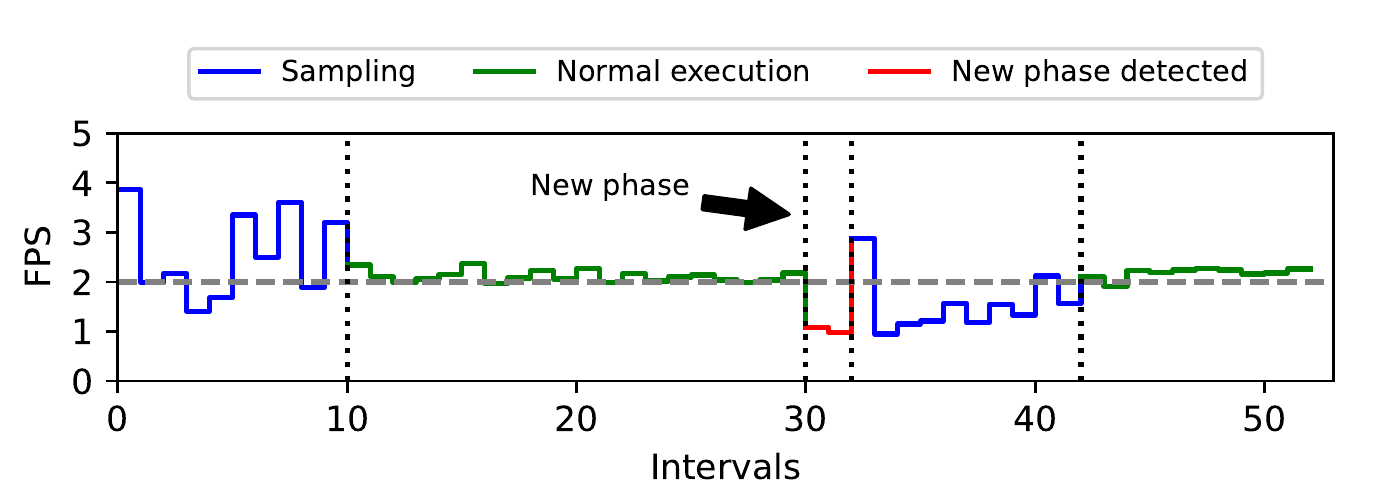}\label{fig:phase_fps}}
    \vspace{-10pt}
    \subfloat[Power dissipation]{\includegraphics[scale=0.6]{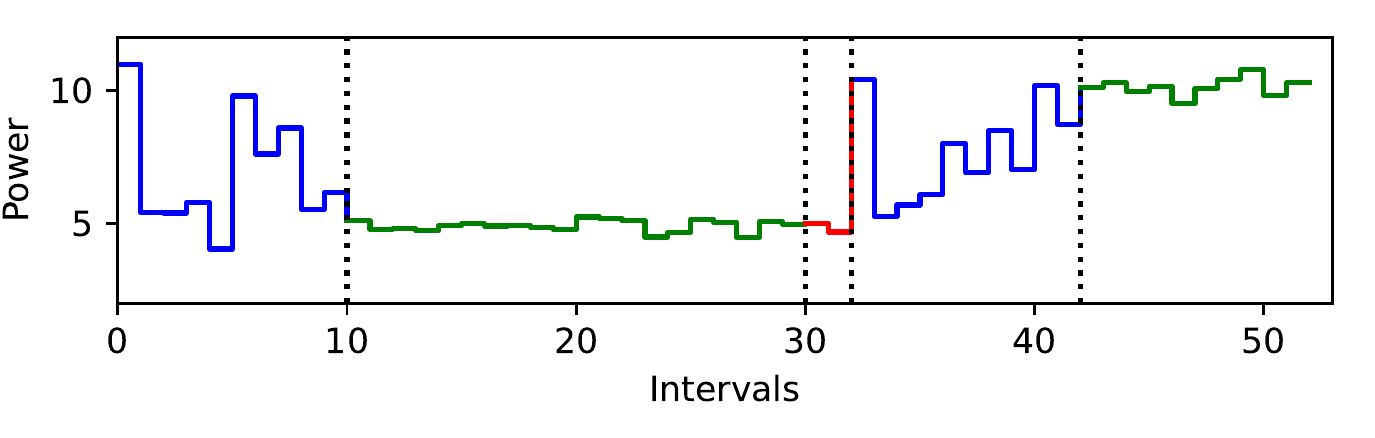}\label{fig:phase_power}}

    \caption{Executing X264 with input that contains phases on \odroid given a constrained optimization problem: least power over a FPS requirement of 2.}
    \label{fig:phase_detect}
\end{figure}

The run-time FPS and power data in chronological order is shown in
Figure~\ref{fig:phase_detect}. \name starts with an sampling phase and
one knob setting is picked after 10 samples evaluated.
However, at round 30, input video changes to ``Ducks take off''. It contains
photographic content so it needs more computation to encode. The FPS
suddenly drops under 2 and the phase detector activates a new sampling
phase after two measuring intervals. In the new phase, a knob setting with much
higher computation capacity and power is picked in order to meet the constraint.
This shows that \app{X264} is able to minimize its power dissipation
while meeting the constraint despite the phase change occurs during the
execution when using \name.


\subsection{Application \& Device Knob Space}
\label{sec:combined_space}
Application knobs are typically programmed to be set before an application starts instead of changing adaptively.
Though tuning application knobs are shown to be beneficial in many prior
works~\cite{green, rumba, slambooster, capri},
additional code change are often required if application knobs need optimization in \name.

In this section, we test the impact of tuning application and device knobs
together for application \app{text classification}. We observe that batch size has
interesting impact on this application's performance.
Its default batch size is 128 and the optimal knob setting on
the desktop system is 7 cores. However, if the batch size is set
to 64 (256), the optimal performance is increased by 11\% (7\%) and only uses
3 (6) cores.

After incorporating batch size ($[32, 64, 128, 256, 512]$) as an additional knob
to the number of cores, the performance of \app{text classification} is furthur
improved by 8\% (246.95 $\rightarrow$ 265.2),
showing \name can tune both application and device knobs.

\subsection{Reuse previous samples}
It is common to run the same run-time configuration for multiple time in
practice (\emph{e.g.} debugging). In this paper, we also evaluate \name's effectiveness
when previous experiments of the same run-time configuration are available.
In this case, sampling history of previous experiments are utilized to
construct a more accurate surrogate model for \bo and Gaussian process regressor.
For the run-time configuration discussed in Section~\ref{sec:eval:perf_odroid},
the average $QoS$ loss drops from 4.8\% to 3.6\% when one previous experiment is available.
When more than three previous runs are available, the average $QoS$ loss drops below 3\%.

%

\subsection{Discussion}
\label{sec:discussion}

\paragraph{\textbf{Controller Overhead.}}
Since the sample size is small in our
experiments, model building or update only takes $\sim$0.2s on the server.
As application's execution does not stall during model building or update,
this duration has no impact on system's performance.

\paragraph{\textbf{Number of Samples.}}
\label{sec:len_sampling}
In this paper, we normalize application's total execution time to be $\sim$10x of a sampling phase.
In general, if the amount of workload is fixed, the more samples, the better
knob setting choice after a sampling phase, but there are two concerns: 1) increased overhead by sampling and evaluating more sub-optimal knobs; 2) less time for the picked knob to benefit the whole execution.
In this paper, we try to make it as small as possible for it to be more practical while yielding reasonable results.

\paragraph{\textbf{Granularity of Sampling Interval.}}
Given a fixed sampling phase duration, the shorter the sampling interval is, the
more samples can be evaluated but more noise is involved in each sample as a cost.
On the other hand, less samples can be taken when using a longer interval,
leading to less chance in finding a good solution.

\paragraph{\textbf{Non-streaming applications.}}
\name may not be useful for non-streaming applications.
If the target application's duration is short, there won't be enough time for the sampling strategy to be beneficial.
If the application is long-running but contains multiple small phases, sampling can be misleading.

\section{Related Work}
\label{sec:related_work}

\paragraph{\textbf{Profiling-based approaches}}

Many systems have been proposed for solving constrained optimization
problems such as trading off program accuracy and performance
~\cite{green, gmm, neuralvector, jouleguard, poet, rumba, caloree, hbm, thunderx_beacon, mimo, spectr, capri}.
Most of them rely on profiling data to construct models for control.
For example, Rumba~\cite{rumba} applies lightweight error checks during application's execution
to detect large approximation errors based on an offline error predictor,
and then fixes these errors by re-computation.
Green~\cite{green} periodically execute the exact version of the target program
in parallel with a pre-defined approximate version and tunes knobs based on their $QoS$ difference.
NeuralVector~\cite{neuralvector} trains embeddings for loop code snippets and tries to
find the best parameters for openmp \emph{pragmas} before execution.
POET~\cite{poet} collects power and performance data exhaustively offline and
use them during online control with a PID-like controller.
MIMO~\cite{mimo} trains offline multi-input multi-output models and uses them in a
traditional control loop.
Profiling-based approaches are infeasible to adapt to diverse run-time
configuration space.

\paragraph{\textbf{Online approaches\textbf}}
Online approaches only utilize the information of the target application at run
time to tune knob settings~\cite{siblingrivalry, fang2014performance, li2006dynamic,
slambooster, slambooster2, flicker, ponomarev2001reducing, holistic}.
For example, SiblingRivalry~\cite{siblingrivalry} and Flicker~\cite{flicker} use
evolutionary algorithms to find the optimal hardware allocation for applications.
Work~\cite{fang2014performance} utilizes simulated annealing to tune knobs.
However, evolutionary algorithm and simulated annealing take much more
samples than \bo does to converge to a near-optimal result.
SLAMbooster~\cite{slambooster} applies PID controller to hot domains such as
Simultaneous Localization and Mapping.
However, such approach requires insights into application's knob space ahead of
time.



\paragraph{\textbf{Sampling-based control}}
Sampling-based control are primarily used in certain domains such as robotic motion
planning~\cite{karaman2011sampling, kingston2018sampling, liu2010sampling},
where robots with many degrees of freedom try to find feasible motion sequences
autonomously under constraints for operations in realistic tasks such as
spacecraft logistics, health care, \emph{etc}.

\paragraph{\textbf{\bo use cases}} \bo gets increasingly popular in solving sampling-constrained problems such as
hyperparameter tuning for machine (deep) learning models~\cite{feurer2019auto, snoek2012practical, snoek2015scalable},
neural architecture search~\cite{elsken2019neural, kandasamy2018neural},
robotic control~\cite{antonova2017deep, calandra2016bayesian},
chemical material design~\cite{griffiths2017constrained, li2017rapid}, \emph{etc}.
\bo also gets more attention in designing specific control problems in recent years,
including finding the best controller to certain industrial process ~\cite{neumann2019data};
tuning PID controllers for HVAC systems~\cite{fiducioso2019safe};
modeling aircraft maneuver control system~\cite{kim2019black}, \emph{etc}.

\section{Conclusion}
\label{sec:conclusion}

In the paper, we propose an online controller to
optimize streaming applications entirely at run time.
At the beginning of each phase of an experiment, \name samples and evaluates knob settings
sequentially to determine a good one for the remaining execution according to a user-defined constrained optimization problem.
\name applies a hybrid approach of machine learning regressors and \bo to improve the choices of samples.
We demonstrate \name's effectiveness by evaluating multiple applications across
multiple devices, and showing \name is able to find near-optimal knob settings at run time.

\newpage
\bibliographystyle{ACM-Reference-Format}
\bibliography{references}

\end{document}